\documentclass[graybox]{svmult}

\usepackage{mathptmx}           
\usepackage{bm}                 
\usepackage{helvet}             
\usepackage{courier}            
\usepackage{type1cm}            
\usepackage{tikz}
\usepackage{circledsteps}
\usepackage{xcolor}
\usepackage{makeidx}            
\usepackage{graphicx}           
\usepackage{multicol}           
\usepackage[bottom]{footmisc}   

\makeindex            

\begin{document}

\title*{Runtime Energy Monitoring for RISC-V Soft-Cores}
\titlerunning{Runtime Energy Monitoring for RISC-V Soft-Cores}
\authorrunning{A. Scionti et al.}

\author{Alberto Scionti and Paolo Savio and Francesco Lubrano and Olivier Terzo and Marco Ferretti and Florin Apopei and Juri Bellucci and Ennio Spano and Luca Carriere}

\institute{
Alberto Scionti, Paolo Savio, Francesco Lubrano, Olivier Terzo \at LINKS Foundation, Turin (ITALY), \email{name.surname@linksfoundation.com}
\and Marco Ferretti, Florin Apopei \at Teseo SpA -- EES CLEMESSY, Druento (ITALY) \email{name.surname@eiffage.com}
\and Juri Bellucci \at Morfo Design srl, Sesto Fiorentino (ITALY) \email{juri.bellucci@morfodesign.it}
\and Ennio Spano, Luca Carriere \at GE Avio srl, Rivalta di Torino (ITALY) \email{name.surname@avioaero.it}
}

\maketitle

\abstract*{Energy efficiency is one of the major concern in designing advanced computing infrastructures. From single nodes to large-scale systems (data centers), monitoring the energy consumption of the computing system when applications run is a critical task. Designers and application developers often rely on software tools and detailed architectural models to extract meaningful information and determine the system energy consumption. However, when a design space exploration is required, designers may incur in continuous tuning of the models to match with the system under evaluation. To overcome such limitations, we propose a holistic approach to monitor energy consumption at runtime without the need of running complex (micro-)architectural models. Our approach is based on a measurement board coupled with a FPGA-based System-on-Module. The measuring board captures currents and voltages (up to tens measuring points) driving the FPGA and exposes such values through a specific memory region. A running service reads and computes energy consumption statistics without consuming extra resources on the FPGA device. Our approach is also scalable to monitoring of multi-nodes infrastructures (clusters). We aim to leverage this framework to perform experiments in the context of an aeronautical design application; specifically, we will look at optimizing performance and energy consumption of a shallow artificial neural network on RISC-V based soft-cores.}

\abstract{Energy efficiency is one of the major concern in designing advanced computing infrastructures. From single nodes to large-scale systems (data centers), monitoring the energy consumption of the computing system when applications run is a critical task. Designers and application developers often rely on software tools and detailed architectural models to extract meaningful information and determine the system energy consumption. However, when a design space exploration is required, designers may incur in continuous tuning of the models to match with the system under evaluation. To overcome such limitations, we propose a holistic approach to monitor energy consumption at runtime without the need of running complex (micro-)architectural models. Our approach is based on a measurement board coupled with a FPGA-based System-on-Module. The measuring board captures currents and voltages (up to tens measuring points) driving the FPGA and exposes such values through a specific memory region. A running service reads and computes energy consumption statistics without consuming extra resources on the FPGA device. Our approach is also scalable to monitoring of multi-nodes infrastructures (clusters). We aim to leverage this framework to perform experiments in the context of an aeronautical design application; specifically, we will look at optimizing performance and energy consumption of a shallow artificial neural network on RISC-V based soft-cores.}

\section{Introduction}
\label{sec:introduction}
Energy efficiency became a major concern of modern computing systems; extracting the highest performance possible for a target application domain is essential to continue scaling up cloud and HPC infrastructures. Monitoring energy consumption is a complex task that requires software tools and detailed (micro-)architectural models to precisely matching with the target system. Despite their undoubted value, monitoring energy (at runtime) through such software and models incurs in an additional overhead (i.e., computing resources are partly used to run the monitor). Designers and application developers need to explore design spaces to find the optimal system setup; in these cases, software tools and energy models must be continuously tuned to match with the specific target system.

In a quest for energy efficient computing system, the RISC-V architecture emerged with the promise of being a first in-class when compared with X86 and ARM counterparts. RISC-V provides a royalty-free customizable modular architecture (i.e., it allows for the definition of custom instructions and instruction set extensions) that is attracting the attention of hardware designers. On the other hand, software developer community moved fast to port applications and developing frameworks on it and, as of today, most of the commonly used operating systems (e.g., many Linux and RTOS distributions) and tools are available on RISC-V systems. Although RISC-V architecture implementations are still in a relatively early stage of development compared to X86 and ARM, they have proven to be a valuable alternative, particularly in use cases where power consumption is a critical concern~\cite{suarez2024}. For these reasons, the number of compute system based on RISC-V is rapidly growing, albeit they mostly resemble IoT/Edge computing platforms. Apart from very few examples, when dealing with high-performance computing (HPC) platforms, RISC-V is years or even decades away from replacing X86 or ARM. However, also in this sector some interesting systems are appearing. In 2023, Milk-V presented the Pioneer system, a desktop-class machine equipped with a 64-cores processor~\cite{brown2023risc}. In late 2023, Ventana announced the Veyron V2, a 192-cores RISC-V CPU (enhancing over the previous generation), which reduced the performance gap with X86 and ARM server-class processors~\cite{ventana2023}. Finally, also several prototyping platforms have been developed~\cite{bartolini2022monte, diehl2024preparing}.

Field Programmable Gate Arrays (FPGAs) have been used since long time as flexible prototyping platforms. Following technological progresses, modern FPGA devices, coupling ever larger reconfigurable fabrics with powerful hardened IP cores, are now part of large cloud and HPC infrastructures. For example, Amazon Web Services (AWS) offers EC2 F1 instances, which integrate Xilinx UltraScale+ FPGAs to provide customizable hardware acceleration for applications such as machine learning inference, data analytics, video processing, and genomics. These instances enable developers to deploy custom FPGA designs in the cloud, leveraging AWS's scalable infrastructure to enhance performance and reduce time to market.

All the major FPGA vendors offer Systems-on-Chip (SoCs) equipped with FPGA fabrics, multicore processor subsystems and accelerator engines. Microchip's PolarFire is a recent midrange SoC equipped with a reconfigurable fabric and five 64 bit RISC-V cores, which are able to run a full fledged Linux operating system. It is available as an integrated development board, as well as a compact System-on-Module (SoM), i.e., a small board where power delivery lines, I/O and peripherals are routed out through dedicated connectors. Systems like this can be effectively used to address the design and evaluation of custom implementation of RISC-V based processors (including application specific ones) without incurring in high manufacturing costs: with a well established process for compiling and mapping high-level descriptions to physical resources of the programmable fabric, designers and application developers are enabled to easily perform extensive design space explorations.

Aeronautic domain makes large use of HPC infrastructures to accelerate the execution of complex engineering applications. As such, aeronautic industry is always looking at novel compute solutions for keeping the pace with the growing complexity of the simulated systems. One of the major innovation in the last years is the introduction of surrogate models, which are based on artificial intelligence techniques to reduce the number of simulations required to sample large parametrized design spaces. Since in many cases it has been demonstrated that a population of shallow artificial neural networks (ANNs) performs well (this kind of networks badly fit on GPUs and AI chips, because these latter are optimized for large models), being able to characterize their execution (training and inference) on novel RISC-V implementations is important to drive the future decision making process in selecting the most appropriate processor profile once they will become available for large deployments.

In this paper, we propose a holistic framework for monitoring runtime energy consumption of RISC-V based compute nodes. To this purpose, we propose the design and integration of a custom measuring board with a FPGA based SoM, which is our preferred target for running RISC-V soft-cores. With up to tens of sensing elements measuring the current drawn at different points of the FPGA SoC, the proposed system can provides a detailed map of the system energy consumption. Sampled data can be correlated with the running soft-core features, thus providing an important tool for optimizing the micro-architecture.

\section{Background}
\label{sec:background}
RISC-V architecture is gaining momentum, with many development platforms being available on the market and some research projects showcasing prototypes. Despite most of the today's platforms resembling single board computers (SBCs) are far away to compete with X86 and ARM based systems in the HPC arena, they have been proved adequate for developing applications and frameworks. 

Among these devices, one of the newly released boards is the SiFive HiFive Premier P550~\cite{premierp550}. The Hifive Premier P550 is an SBC powered by the ESWIN EIC7700X SoC, featuring a SiFive Quad-Core P550 64-bit Out-of-Order CPU cluster, 16 GB or 32 GB of LPDDR5 RAM, and a 128 GB eMMC with Ubuntu 24.04 pre-installed. This board offers a ready-to-go solution for developers who want to develop or port applications for RISC-V, providing sufficient performance for a variety of use cases.

Although the Premier P550 does not implement vector extension (RVV), it includes an AI NPU with around 20 TOPS of peak capacity to execute AI-related calculations-such as matrix multiplications or convolution operations-making it a valuable solution for a variety of edge-based machine learning tasks, like real-time image recognition, speech processing, or even advanced sensor data fusion, while maintaining a strong balance between performance and energy efficiency. At the time of writing, the pre-installed Ubuntu distribution does not support the NPU; however, the situation is rapidly evolving and there are several options to overcome this issue. One of them is to install a custom Debian\footnote{https://github.com/eswincomputing/eic7x-images/tree/Debian-v1.0.0-p550-20241230} which is open source and implement the support for the NPU. 

The Premier P550 is particularly specialized for edge use cases, where the energy efficiency is one of the major concerns. Thus, the board is equipped with a state-of-the-art microcontroller unit (MCU), which delivers energy monitoring data, critical for applications such as battery management, renewable energy systems, and dynamic power scaling. This MCU offers two interfaces to access status and monitoring data:
\begin{itemize}
    \item CLI, accessible through the UART terminal interface of the MCU
    \item WEB GUI, accessible connecting to a dedicated Ethernet RJ45 port
\end{itemize}
Both methods provide data related to power consumption, voltage and current.

Characterizing workloads running on the Premier P550 using the power consumption data provided by the MCU is not a trivial task. It requires a correlation process among at least the CPU usage data with the MCU's measurements. The correlation process must account for time synchronization, accurate timestamps and different sampling rates to to ensure data is correctly aligned. Achieving this in real time and/or scaling the process to a cluster of many boards, presents additional challenges and issues due to the lack of a direct connection among the system and the MCU for what concern monitoring data.

Besides SBCs, RISC-V is also rapidly moving into the server market segment, with the server-class profile already ratified, and vector extensions (RVV) being implemented on some SoCs. Several companies within the RISC‑V ecosystem are introducing products in this segment; Ventana Micro System recently announced Veyron~\cite{ventanaMicroSystems2025}, a server-class SoC combining 16 Out-of-Order high-performance cores and supporting RVV v1.0, with 48MiB of L3 cache memory. This design provides hypervisor hardware acceleration (virtualization) and support for RAS (e.g., ECC, error logging/scrubbing, data poisoning, etc.), which are ideal for cloud environments. The Veyron V2 further extends these features to a 192 cores CPU. Tenstorrent Wormhole~\cite{ignjatovic2022wormhole} presents enhancements in many aspects with respect to the previous generation of AI acceleration platform (Grayskull~\cite{brown2024accelerating}); these include a higher amount of SRAM per tile\footnote{Each tile comprises the network-on-chip (NoC) interface, the SRAM block, 5 RISC-V cores which are assigned to different functions (NoC in/out data movers, computing), and a matrix/vector engine. The Grayskull system counts up to 120 tiles, while the Wormhole stops at 80.} (1.5MiB vs 1.0MiB), higher performance (110 TFLOPS vs 92 TFLOPS), larger memory system (12GiB vs 8GiB) and the support for network communication. As many other AI accelerators, its design has been optimized for processing large AI models.

Many funded projects explored the RISC-V royalty-free and open ISA nature to perform a design space exploration aimed to optimize the system microarchitecture for the specific application domain. For instance, the European Processor Initiative (EPI) developed and taped out an acceleration platform called EPAC~\cite{kovavc2022european}. It combines the Avispado~\cite{semiDynamics2025} RISC-V core with vector processing units (VPU Vitruvius+~\cite{minervini2023vitruvius+}) implementing the RVV v0.7.1, a pair of stencil cores (Stencil and Tensor Accelerator --STX) optimized for AI and machine learning tasks, and a variable precision processing (VPR) core. Interestingly, this solution is able to boot and run a full-fledged Linux OS. Sargantana~\cite{soria2022sargantana} is an alternative design developed by the Barcelona Supercomputing Center, while PULP~\cite{pulp2025} is an open project (mainly maintained by ETH Zurich) which provides several CPU designs ranging from the IoT-optimized to the HPC domain. 

Micro-architecture optimizations require a flexible platform where to iteratively implement and test designs. To this end, soft-cores (i.e., HDL descriptions that can be synthesized and run on top of FPGAs) perfectly match with this need, with many RISC-V designs available for research purposes. Such designs cover an ample set of micro-architectures, ranging from small 16 bit microcontrollers to in-order 32 bit CPUs to complex 64 bit out-of-order CPUs~\cite{zaruba2019cost,openC910,zhao2020sonicboom}. Simulation platforms like FireSim~\cite{karandikar2019using} are limited to monitoring system performance (by capturing traces of software running on the simulated cores); as such, they do not provide any feature to capture runtime power/energy consumption. An alternative approach involves using pure software simulations, such as GEM5 for architectural simulation and McPAT for power modeling, which offer detailed and accurate energy estimation models. Here, the principal limitation is in the need for continuously tuning the model to match the characteristics of the studied micro-architecture. To overcome these limitations, we propose an integrated holistic approach closely resembling the ExaMon framework~\cite{borghesi2021examon}. Our framework adds a hardware mechanism to capture measures relevant to determine the energy consumption of the studied soft-core. With respect to ExaMon, our solution operates with a finer granularity.

\subsection{ANN-based surrogate models}
\label{subsec:ai_surrogate}
Aeronautic design involves performing complex computational fluid dynamic (CFD) simulations, which are very time consuming. In the past decades, artificial neural networks (ANNs) have been proved to speed-up this process, by learning how to associate physical design parameters (input) to performance values (output). For instance, when optimizing the geometry of a turbine, small shallow ANNs demonstrated to be a powerful tool. These networks implement few fully connected layers, each with 10s of neurons. As such, the demand in terms of hardware resources for training and running the networks is limited, and can be conveniently sustained by modern CPUs. In this context, the RISC-V architecture offers an invaluable opportunity to study and optimize modern CPU micro-architectures for energy efficiency. 
%

In~\cite{zhao2020sonicboom}, SonicBOOM soft-core (an Out-of-Order superscalar core, supporting also acceleration for vector/matrix operations) has been extensively evaluated and compared with other architectures and RISC-V soft-core implementations. On the CoreMark/MHz metric, it provides $1.216\times$ more performance compared to the SiFive U74 core, which is representative of many systems available on the market. SonicBOOM will be one of the target implementations from where to start performing design parameters space exploration campaigns aimed at finding the most energy efficient configuration to train and run ANN-based surrogate models.
\section{Monitoring RISC-V soft-cores clusters at runtime}
\label{sec:energy_monitoring}
\begin{figure}[ht]
  \centering
  \includegraphics[width=0.90\linewidth]{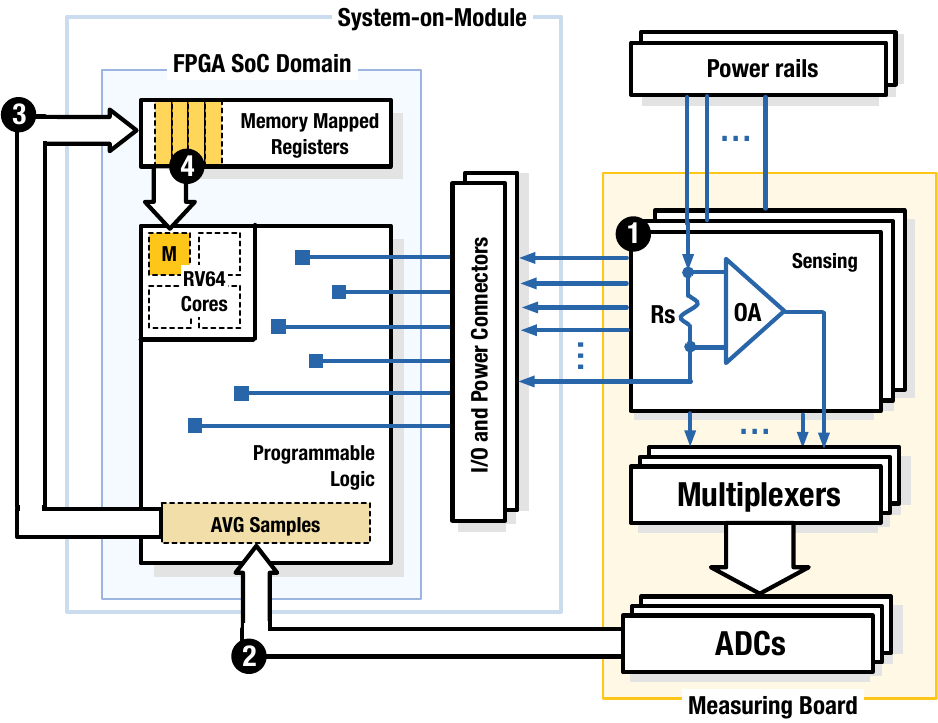}
  \caption{Architectural view of the runtime energy monitor.}
    \label{fig:runtime_energy_monitor}
\end{figure}

Monitoring the energy consumption of a compute system requires to sample its power drawn over a given time interval. By keeping this time interval small, a precise runtime tracking can be achieved.
Figure~\ref{fig:runtime_energy_monitor} depicts the architecture of our runtime energy monitor. We take advantage of the modular design of the target system, which consists of the FPGA SoC mounted on a small form-factor board (referred to as the System-on-Module --SoM) and a carrier board, to easily get access to the FPGA SoC power rails. Indeed all the I/O connections to the peripherals (e.g., SD-card reader, 1 Gbps Ethernet, GPIOs, etc.) and the power rails are routed to the SoM through dedicated connectors. By interposing the measurement board between the SoM and the carrier board, we can arrange multiple sensing elements. From the measurement perspective, sensing the voltage is quite simple, since the power signals need only to pass through an analog-to-digital converter (ADC). On the contrary, sensing the current drawn through the specific line requires a dedicated measuring scheme. From figure~\ref{fig:runtime_energy_monitor}, the reader can see that the sensing elements are inserted between the main power delivery and the load, i.e., the SoM (upper-side insertion scheme \Circled[inner color=white, fill color=black]{1}). This also allows to avoid ground disturbances over the measurements. The sensing element is composed of a \textit{shunt resistor} ($R_s$) coupled with a \textit{signal conditioning} circuit; the former provide the mean for measuring the current flowing in the line without causing a large voltage drop. The measurement happens indirectly, since the differential voltage is actually caught. The signal conditioning circuit (i.e., an operational amplifier based circuit) is needed to make the sensed voltage signal fitting in the range of accepted voltages of the ADCs. Whilst integrating a high number of sensing elements is not a critical task, keeping a 1 to 1 ratio with the ADCs is more complex. To overcome this limitation, we designed our system with a 1-to-N ratio; then, a group of multiplexers are used to connect with the ADCs. 

The output of the ADCs (\Circled[inner color=white, fill color=black]{2}) is brought to the FPGA SoC by using I/O lines available in the SoM connectors. On the FPGA side, the digitalized values are collected by a dedicated logic block, which is in charge of computing the average value over a window of $K$ samples (i.e., AVG Sample). A simple finite state machine (FSM) is implemented to compute the average samples over the whole set of ADCs and making them available on a group of memory mapped registers (\Circled[inner color=white, fill color=black]{3}). The averaged values are also associated to a high-precision timestamp, to make easier reconstructing power (energy) profiles over time.

The ADC values can easily retrieved by one of the compute cores available on the FPGA-SoC (\Circled[inner color=white, fill color=black]{4}). Interestingly, many FPGA SoCs expose asymmetric processing capabilities, i.e., some of the available cores can be configured to run as bare metal cores. As such, these cores are not visible to the operating system, whilst they execute a specific firmware within an infinite loop. Interaction with OS-managed cores (sometime referred to as application cores) is achieved by using communication mailboxes. In our runtime energy monitor, one of the bare metal cores is in charge of transforming such raw values (averaged samples) into actual values of current. To this end, each raw value (representing voltage values in steps) is traced back to the measured current by means of the Ohm's law (the value of the shunt resistor is known). We refer to this core as the \textit{runtime monitor} (M). 

In our proposed framework, the \textit{minimum sampling time interval} ($t_s^{min}$) is imposed by the minimum amount of time required by the ADC circuitry to sample the input signal, which we assume to varying slowly over the time (i.e., we can reasonably assume a bandwidth lower than $1$~MHz). This said, the time interval for the acquisition of an averaged sample is set to $T_{avg} = t_s^{min} \cdot K \cdot N$. On modern FPGA SoC, embedded compute cores run at a relative high clock frequency (generally, between $100$~MHz and $1$~GHz), providing enough space for reading averaged values from memory mapped registers. This help us to optimize the $K$ and $N$ parameters to limit the number of active electronic elements in the measurement board.
\begin{figure}
    \centering
    \includegraphics[width=0.75\linewidth]{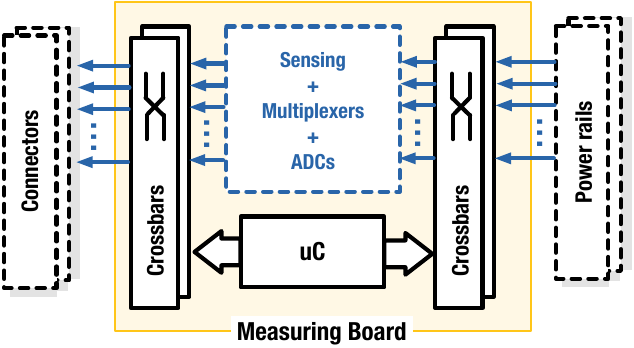}
    \caption{Measurement board extension for agnostic FPGA SoM support.}
    \label{fig:board_extension}
\end{figure}
Making the runtime energy monitor agnostic with regards to the FPGA SoC architecture is achieved by adding some electronics. Instead of designing different measurement boards for different FPGA SoC architectures, we opted for making our design adaptable (see figure~\ref{fig:board_extension}). To this end, connector lines are managed through a configurable crossbar element. Crossbar elements enable the dynamic connection between multiple inputs and outputs without fixed wiring. This allows to configure the mapping between sensing element lines and actual connector pin on the System-on-Module. A small 16 bit microcontroller is used to manage the configuration of the crossbar elements.

\subsection{On-line telemetry}
\label{subsec:telemetry}
Application cores are utilized to implement a (simple) user space daemon, referred to as \textit{energy collector service}, that retrieves measured values from the runtime monitor and performs some post-processing. To limit the impact on the overall system performance, the energy collector service computes the consumed energy from the measured current, voltage and duration of each sample. The computed data is then published in a specific topic using MQTT. MQTT is a lightweight messaging protocol specifically designed for applications with limited network bandwidth, that is build on top of TCP/IP and that is well suited
for the transmission of sensor data coming from devices having small processing capabilities and large network latencies.

Published messages are collected by an external device, which acts as subscriber to the several MQTT topics.
This centralised approach is particularly valuable when multiple nodes are monitored, such as in a cloud or HPC cluster, and allows to store collected data from several devices in a centralised way on a time-series database (e.g., InfluxDB). Visualization tools like Grafana can be deployed and integrated to display power consumption trends through dashboards and charts. This setup facilitates both real-time monitoring and later offline data analysis, providing users with a comprehensive view of the system's energy performance over time.


\section{Conclusion}
\label{subsec:conclusion}
This paper presents an holistic framework for runtime energy consumption measurements of RISC-V soft-cores. The proposed approach is designed to be both flexible and extensible, and we plan to use it to optimize soft-core micro-architectures for ANN-based aeronautical surrogate models. Indeed, (software) simulation-based solutions often require extensive tuning of the energy models for each specific micro-architecture, while hardware based simulations lack fine-grained energy measurement capabilities. At the time of writing this paper, we are in the prototyping phase of our measurement boards, thus leaving an extensive experimental campaign as a near future activity, along with the integration of the energy collector service.

%
%

\begin{thebibliography}{99.}%

\bibitem{brown2023risc} Brown, Nick, Maurice Jamieson, Joseph Lee, and Paul Wang. 2023. Is RISC-V ready for HPC prime-time: Evaluating the 64-core Sophon SG2042 RISC-V CPU. \textit{Proceedings of the SC'23 Workshops of The International Conference on High Performance Computing, Network, Storage, and Analysis}, 1566--1574.

\bibitem{bartolini2022monte} Bartolini, Andrea, Federico Ficarelli, Emanuele Parisi, Francesco Beneventi, Francesco Barchi, Daniele Gregori, Fabrizio Magugliani, Marco Cicala, Cosimo Gianfreda, Daniele Cesarini, et al. 2022. Monte Cimone: paving the road for the first generation of RISC-V high-performance computers. \textit{2022 IEEE 35th International System-on-Chip Conference (SOCC)}, 1--6. IEEE.

\bibitem{diehl2024preparing} Diehl, Patrick, Panagiotis Syskakis, Gregor Daiß, Steven R. Brandt, Alireza Kheirkhahan, Srinivas Yadav Singanaboina, Dominic Marcello, Chris Taylor, John Leidel, and Hartmut Kaiser. 2024. Preparing for HPC on RISC-V: Examining Vectorization and Distributed Performance of an Astrophyiscs Application with HPX and Kokkos. \textit{arXiv preprint} arXiv:2407.00026.

\bibitem{zhao2020sonicboom} Zhao, Jerry, Ben Korpan, Abraham Gonzalez, and Krste Asanovic. 2020. Sonicboom: The 3rd generation Berkeley out-of-order machine. \textit{Fourth Workshop on Computer Architecture Research with RISC-V}, 5:1--7.

\bibitem{ventana2023} Ventana Micro Systems. 2023. Ventana Micro Systems Veyron V2. Available at: \url{https://www.ventanamicro.com/ventana-introduces-veyron-v2/} (Accessed: 2025-01-12).

\bibitem{ventanaMicroSystems2025} Ventana Micro Systems. 2025. Ventana Micro Systems Veyron V1 \& V2. Available at: \url{https://www.ventanamicro.com/technology/risc-v-cpu-ip/} (Accessed: 2025-01-12).

\bibitem{premierp550} SiFive. 2025. SiFive Hifive Premier P550 development board. 
\url{https://www.sifive.com/boards/hifive-premier-p550} (Accessed: 2025-03-25)

\bibitem{ignjatovic2022wormhole} Ignjatović, Drago, Daniel W. Bailey, and Ljubisa Bajić. 2022. The wormhole AI training processor. \textit{2022 IEEE International Solid-State Circuits Conference (ISSCC)}, 65:356--358. IEEE.

\bibitem{brown2024accelerating} Brown, Nick and Ryan Barton. 2024. Accelerating stencils on the Tenstorrent Grayskull RISC-V accelerator. \textit{SC24-W: Workshops of the International Conference for High Performance Computing, Networking, Storage and Analysis}. IEEE.

\bibitem{kovavc2022european} Kovač, Mario, Jean-Marc Denis, Philippe Notton, Etienne Walter, Denis Dutoit, Frank Badstuebner, Stephan Stilkerich, Christian Feldmann, Benoît Dinechin, Renaud Stevens, et al. 2022. European processor initiative: Europe's approach to exascale computing. In \textit{HPC, Big Data, and AI Convergence Towards Exascale}, 273--290. CRC Press.

\bibitem{semiDynamics2025} Semidynamics. 2025. Semidynamics Avispado RISC-V core. Available at: \url{https://semidynamics.com/en/products/avispado} (Accessed: 2025-01-12).

\bibitem{minervini2023vitruvius+} Minervini, Francesco, Oscar Palomar, Osman Unsal, Enrico Reggiani, Josue Quiroga, Joan Marimon, Carlos Rojas, Roger Figueras, Abraham Ruiz, Alberto Gonzalez, et al. 2023. Vitruvius+: an area-efficient RISC-V decoupled vector coprocessor for high performance computing applications. \textit{ACM Transactions on Architecture and Code Optimization}, 20(2):1--25.

\bibitem{soria2022sargantana} Soria-Pardos, Víctor, Max Doblas, Guillem López–Paradís, Gerard Candón, Narcís Rodas, Xavier Carril, Pau Fontova–Musté, Neiel Leyva, Santiago Marco-Sola, and Miquel Moretó. 2022. Sargantana: A 1 GHz+ in-order RISC-V processor with SIMD vector extensions in 22nm FD-SOI. \textit{2022 25th Euromicro Conference on Digital System Design (DSD)}, 254--261. IEEE.

\bibitem{pulp2025} PULP platform. 2025. PULP platform. Available at: \url{https://pulp-platform.org/index.html} (Accessed: 2025-01-12).

\bibitem{zaruba2019cost} Zaruba, Florian and Luca Benini. 2019. The cost of application-class processing: Energy and performance analysis of a Linux-ready 1.7-GHz 64-bit RISC-V core in 22-nm FDSOI technology. \textit{IEEE Transactions on Very Large Scale Integration (VLSI) Systems}, 27(11):2629--2640.

\bibitem{openC910} OpenC910 Soft-Core. Available at: \url{https://github.com/XUANTIE-RV/openc910} (Accessed: 2025-01-12).

\bibitem{karandikar2019using} Karandikar, Sagar, David Biancolin, Alon Amid, Nathan Pemberton, et al. 2019. Using FireSim to Enable Agile End-to-End RISC-V Computer Architecture Research.

\bibitem{borghesi2021examon} Borghesi, Andrea, Alessio Burrello, and Andrea Bartolini. 2021. Examon-x: a predictive maintenance framework for automatic monitoring in industrial IoT systems. \textit{IEEE Internet of Things Journal}, 10(4):2995--3005.

\bibitem{suarez2024} Suárez, Daniel, Francisco Almeida, and Vicente Blanco. "Comprehensive analysis of energy efficiency and performance of ARM and RISC-V SoCs.\textit{The Journal of Supercomputing}, 80.9 (2024): 12771-12789.
\end{thebibliography}
%

\end{document}